\documentclass{PoS}

\usepackage{epsf}
\usepackage{epsfig,rotate}
\usepackage{afterpage}
\usepackage{longtable}
\usepackage{cite}
\usepackage{color}
\let\normalcolor\relax

\newcommand{\vcb}{|V_{cb}|}

\newcommand{\vub}{|V_{ub}|}

\newcommand{\Br}{\rm Br}

\newcommand{\be}{\begin{equation}}
\newcommand{\ee}{\end{equation}}
\newcommand{\bea}{\begin{eqnarray}}
\newcommand{\eea}{\end{eqnarray}}

\newcommand{\ba}{\begin{array}}
\newcommand{\ea}{\end{array}}

\def\kpn{K^+\rightarrow\pi^+\nu\bar\nu}

\def\klpn{K_{L}\rightarrow\pi^0\nu\bar\nu}

\title{Flavour Visions}

\ShortTitle{Flavour Visions}

\author{\speaker{Andrzej J. Buras} \addtocounter{footnote}{1}\thanks{} \\
Technical University Munich, Physics Department, D-85748 Garching, Germany,\\
TUM-IAS, Lichtenbergstr. 2a, D-85748 Garching, Germany \\
E-mail: \email{Andrzej.Buras@ph.tum.de}}

\abstract{This decade will allow to improve the resolution of 
 the short distance scales by at least  an order of magnitude, 
extending the picture of fundamental physics 
down to scales $5\cdot 10^{-20}$m with the help of the LHC. Further resolution 
down to scales as short as $10^{-21}$m 
 should be possible with the help of 
high precision experiments in which flavour violating processes will play a 
prominent role. Will this increase in resolution allow us to 
see new particles (new animalcula) similarly to what  
Antoni van Leeuvenhoek saw by discovering  bacteria in 1676? The basic question 
for particle physics is how these new animalcula will look like and 
which difficulties of the Standard Model (SM) they will help us to solve 
and which 
new puzzles and problems they will bring with them. I will describe what role
flavour physics will play in these exciting times provided this new 
world is animalculated.
}

\FullConference{The 13th International Conference on B-Physics at Hadron Machines\\
                 April 4-8 2011\\
                 Amsterdam, The Netherlands}

\begin{document}

\section{Overture}

The year 1676 was a very important year for the humanity. In this year 
Antoni van Leeuvenhoek (1632-1723) discovered the empire of bacteria. 
He called these small creatures {\it animalcula} (small animals). This 
discovery was a mile stone in our civilization  for at least two reasons: 
\begin{itemize}
\item
 He discovered invisible to us creatures which over thousands of years 
 were systematcally killing the humans, often responsible for millions 
 of death in one year. While Antoni van Leeuvanhoek did not know that 
 bacteria could be dangerous for humans, his followers like L. Pasteur (1822-1895),
 Robert Koch (1843-1910) and other {\it animalcula hunters} not only realized 
 the danger coming from these tiny creatures but also developed weapons against
 this empire. 
\item
 He was the first human who looked at short distance scales invisible to 
 us, discovering thereby
 a new {\it underground world}. At that time researchers 
 looked mainly at large distances, discovering new planets and finding 
 laws, like Kepler laws, that Izaak Newton was able to derive from his 
 mechanics.
 \end{itemize}
 
While van Leeuvanhoek could reach the resolution down to roughly 
$10^{-6}$m, over the last 335 years this resolution could be improved 
by twelve orders of magnitude. On the way down to shortest distance 
scales scientists discovered {\it nanouniverse} ($10^{-9}$m), 
{\it femtouniverse}   ($10^{-15}$m) relevant for nuclear particle physics 
and low energy elementary particle physics and finally 
{\it attouniverse} ($10^{-18}$m)
that is the territory of contemporary high energy elementary particle physics.

In this decade we will be able to improve the resolution of 
 the short distance scales by at
least
 an order of magnitude, extending the picture of fundamental physics 
down to scales $5\cdot 10^{-20}$m with the help of the LHC. Further resolution 
down to scales as short as $10^{-21}$m ({\it zeptouniverse})
 should be possible with the help of 
high precision experiments in which flavour violating processes will play a 
prominent role. 

In this context one should point out that van Leeuvanhoek was really lucky.
If the animalcula that he discovered where by an order of magnitude smaller, 
he would not see them with the microscopes he built. 
Moreover, the theorists of the 17th century did not make any predictions 
for this new world.
In this sence 
particle physicists are in a better position.  
We are all convinced that some new animalcula must exist at the short 
distance scales explored by the LHC and new high precision experiments. 
Moreover we 
have some ideas how they could look like, even if  there are different 
opinions on their possible appearance.
This talk deals with the search for 
new animalcula  with the help of flavour physics. More details on this 
search can be found in \cite{Buras:2010wr} as well as  
\cite{Isidori:2010kg,Fleischer:2010qb,Nir:2010jr,Antonelli:2009ws,Hurth:2010tk}. I should also warn possible readers that in view of space limitations my 
list of references is incomplete. This is compensated by approximately 300 references 
in \cite{Buras:2010wr}. There is also some overlap with the latter review 
but I made an effort to include also new developments that cannot be found 
there.

\section{Beyond the SM}
\subsection{Preliminaries}
The fundamental Lagrangian of the SM consists of four pieces
\be 
{\cal L}_{\rm SM}= L_{\rm gauge}+L_{\rm fermion}+L_{\rm Higgs}+L_{\rm Yukawa},
\ee
of which the first two terms containing kinetic terms of gauge bosons 
and fermions, invariant under the SM gauge group, 
have been rather well tested at various 
laboratories, in particular at CERN and Fermilab. The situation with 
the last two terms is different. 
\begin{itemize}
\item
$L_{\rm Higgs}$ is responsible for spontaneous breakdown of 
the electroweak symmetry and the 
generation of $W^\pm$ and $Z^0$ masses. It provides also the mass for the 
Higgs boson. The potential in this term 
driving this breakdown, even if consistent with the gauge symmetry, is 
rather adhoc. Clearly we are here at the level similar to the
 Ginzburg-Landau theory of superconductivity. A more dynamical mechanism 
of electroweak symmetry breakdown is absent in the SM.
\item
$L_{\rm Yukawa}$ is responsible for the generation of fermion masses through 
Yukawa-like interactions of fermions with the Higgs system. A natural 
scale for fermion masses generated in this manner is the value of 
the vacuum expectation $v$
of the relevant neutral component of the Higgs doublet, this means 
$246$~GeV in my conventions. This works for the top quark but fails 
totally for the remaining fermions. Their masses are by orders of 
magnitude smaller than $m_t$. Consequently in order to describe 
the observed fermion mass spectrum Yukawa interactions must 
have a very hierarchical structure. This hierarchical structure is 
believed to be responsible for the  observed hierarchy in flavour 
violating interactions of quarks.
However, a convincing theory behind this hierarchy is still missing.
\end{itemize}

Thus in spite of the fact that the SM appears to describe the existing 
data rather well, it does with the help of 28 parameters of which 22 
recide in the flavour sector. 

 Taking all these facts together, the message is clear: in our search for a 
more fundamental theory we need to improve our understanding of electroweak 
symmetry breaking and of flavour which would allow us
to answer the crucial question:

{\bf What is the dynamical origin of the observed electroweak symmetry 
breaking, of  related fermion masses and the reason for their 
 hierarchy and hierarchy of their flavour-changing interactions?}

Related important questions are clearly these ones:
\begin{itemize}
\item
Will the dynamics of electroweak symmetry breaking be driven by 
an elementary Higgs and be calculable within perturbation theory?
\item
Will these dynamics be related to a new strong force with a composite
Higgs or without Higgs at all?
\item
Could these dynamics help us to explain the amount of matter-antimatter 
asymmetry and the amount of dark matter observed in the universe?
\item
Will these dynamics help us to explain various anomalies observed 
recently in the flavour data?
\end{itemize}

Whatever these dynamics will be, we need new particles and new forces in 
order to answer all these questions and this means new animalcula at 
the scales explored by the LHC and high precision experiments. But 
the identification of them is quite challenging both experimentally 
and theoretically. 

In order to illustrate this problematic from the point of view of flavour 
physics, let us consider the tree level decay
$B^+\to\tau^+\nu_\tau$ which in the SM is just mediated by a $W^+$ exchange. 
The resulting branching ratio reads
\begin{equation}
Br(B^+\to\tau^+\nu_\tau)_{\rm SM}=\left|A\frac{g_2^2}{M_W^2}\right|^2,
\end{equation}
where $g_2$ is the $SU(2)_L$ coupling constant and $A$ collects all 
factors that depend on the parameters of the SM.

Let us next assume the presence of a heavy charged boson (scalar or vector) 
$H^+$ mediating this decay as well, so that the branching ratio is 
modified as follows
\begin{equation}
Br(B^+\to\tau^+\nu_\tau)_{\rm SM+H}=
\left|A\frac{g_2^2}{M_W^2}+B\frac{g_H^2}{M_H^2}\right|^2.
\end{equation}
Here $g_H$ is a new coupling constant, $M_H$ is the mass of $H^+$ and 
$B$ collects all factors that depend on the parameters of the new physics (NP)
model.

Finally let us assume that experiments find the  disagreement with the 
SM prediction:
\begin{equation}
Br(B^+\to\tau^+\nu_\tau)_{\rm EXP}-
Br(B^+\to\tau^+\nu_\tau)_{\rm SM}\not=0.
\end{equation}
In principle this deviation could signal the presence of the boson $H^+$ and 
by suitably choosing the coupling $g_H$, the mass $M_H$ and $B$ we could 
obtain the agreement with the data. Yet, clearly we cannot be sure that 
this is really the explanation, as many other NP contributions 
could be responsible for this anomaly. What would definitely help would 
be the discovery of $H^+$ in high energy collisions like those taking 
place at the LHC or TEVATRON, but what if $M_H$ is beyond the reach of 
of these machines? Moreover, even if $H^+$ and other new particles could 
be discovered at the LHC, the measurement of their properties, in 
particular their flavour interactions, both flavour violating and flavour 
conserving will be a real challange. Here rare and CP-violating phenomena 
in low energy, high precision experiment can offer a great help, as they 
did already in the past 50 years. Yet, as we have seen above, a single 
measurement of a rare process, even if signalling the presence of new particles, will not be able to tell us what these particles are.

The message is clear: In order to identify new animalcula through flavour 
physics and generally through high precision experiments we need:
\begin{itemize}
\item
 Many high precision measurements of many observables and precise theory,
\item
Identification of  patterns of flavour violation in various NP models, in 
particular correlations between many flavour observables that could distinguish 
between various NP scenarios,
\item
Identification of 
correlations between low energy flavour observables and observables 
measured in high energy collisions.
\end{itemize}  

Despite  the impressive success of the CKM picture of flavour changing 
interactions \cite{Cabibbo:1963yz,Kobayashi:1973fv} in which the GIM mechanism 
\cite{Glashow:1970gm} for the
suppression of flavour changing neutral currents (FCNC) 
plays a very important role, there are many open questions of 
theoretical and experimental nature that should be answered before we
can claim to have a theory of flavour.
Among the basic questions in flavour physics that could be answered in the 
present decade are the following ones:
\begin{enumerate}
\item
What is the fundamental dynamics behind the electroweak symmetry breaking 
that very likely plays also an important role in flavour physics?
\item
Are there any new flavour symmetries that could
help us to understand the existing hierarchies of fermion masses and the 
hierarchies in the quark and lepton flavour violating interactions? Are they 
local or global? Are they continuous or discrete?
\item
Are there any flavour violating interactions that are not governed by 
the SM Yukawa couplings? In other words, is  the Minimal Flavour Violation 
(MFV)
the whole story?
\item
Are there any additional {\it flavour violating} and CP-violating (CPV) phases that could 
explain certain anomalies present in the flavour data and 
simultaneously play
a role in the explanation of the observed baryon-antibaryon asymmetry 
in the universe (BAU)?
\item
Are there any {\it flavour conserving} CPV phases that could also help
in explaining the flavour anomalies in question and would be signalled 
in this decade  through  enhanced electric dipole moments (EDMs) of the
neutron, the electron and of other particles?
\item
Are there any new sequential heavy quarks and leptons of the 4th 
generation and/or new fermions with exotic quantum numbers like 
vector-like fermions?
\item
Are there any elementary neutral and charged scalar particles  
with masses below 1~TeV and having a significant impact on flavour physics?
\item
Are there any new heavy gauge bosons representing an enlarged gauge 
symmetry group?
\item
Are there any relevant right-handed (RH) weak currents that would help us to
make our fundamental theory parity conserving at short distance scales 
well below those explored by the LHC?
\item
How would one successfully address all these questions if the breakdown of 
the electroweak symmetry would turn out to be of a non-perturbative origin? 
\end{enumerate}

An important question is the following one:
will some of these questions be answered through the interplay of high
energy processes explored by the LHC with low energy precision experiments 
or are the relevant scales of fundamental flavour well beyond the energies
explored by the LHC and future colliders in this century? The existing 
tensions in some of the corners of the SM to be discussed below 
and still a rather big room for
NP contributions in rare decays of mesons and leptons and 
CP-violating observables, including in particular EDMs, give us hopes that 
indeed
 several phenomena required to answer at least
some of these questions could be discovered in this decade.

\subsection{Superstars of Flavour Physics in 2011-2016}
As far as high precision experiments are concerned a number of selected 
processes and observables will, in my opinion, play the leading role in 
learning about the NP in this new territory. This selection is based on 
the sensitivity to NP and theoretical cleanness. The former can be increased 
with the increased precision of experiments and the latter can improve with
the progress in theoretical calculations, in particular the non-perturbative
ones like the lattice simulations.

My superstars for the coming years are as follows:
\begin{itemize}
\item
The mixing induced CP-asymmetry $S_{\psi\phi}(B_s)$ that is
 tiny in the SM: $S_{\psi\phi}\approx 0.04$. The asymmetry
 $S_{\phi\phi}(B_s)$ is also important. It is also
 very strongly suppressed 
in the SM and is sensitive to NP similar to the one explored through 
the departure of $S_{\phi K_S}(B_d)$ from $S_{\psi K_S}(B_d)$ 
\cite{Fleischer:2007wg}.
\item
The rare decays $B_{s,d}\to\mu^+\mu^-$ that could be enhanced in certain 
NP scenarios by an order of magnitude with respect to the SM values.
\item
The angle $\gamma$ of the unitarity triangle (UT) that will be precisely 
measured
through tree level decays.
\item
$B^+\to\tau^+\nu_\tau$ that is sensitive to charged Higgs particles.
\item
The rare decays $K^+\to\pi^+\nu\bar\nu$ and $K_L\to\pi^0\nu\bar\nu$ that
belong to the theoretically cleanest decays in flavour physics.
\item
The decays $B\to X_s\nu\bar\nu$, $B\to K^*\nu\bar\nu$ and $B\to K\nu\bar\nu$ 
that are theoretically rather clean and are sensitive to RH currents.
\item
Numerous angular symmetries and asymmetries in $B\to K^*l^-l^-$.
\item
Lepton flavour violating decays like $\mu\to e\gamma$, $\tau\to e\gamma$, 
$\tau\to\mu\gamma$, decays with three leptons in the final state and 
$\mu-e$ conversion in nuclei.
\item
Electric dipole moments of the neutron, the electron, atoms and leptons.
\item
Anomalous magnetic moment of the muon $(g-2)_\mu$ that indeed seems to
be ''anomalous'' within the SM even after the inclusion of radiative corrections.
\item
The ratio $\varepsilon'/\varepsilon$ in $K_L\to\pi\pi$ decays 
which is known experimentally within 
$10\%$ and which should gain in importance in this decade due to improved 
lattice calculations. 
\item
Precise measurements of two-body $B_d$ and in particular $B_s$ decays which 
in combination with QCD factorization and various flavour symmetries 
   \cite{Fleischer:2010qb} could teach us more about 
the interplay of strong and electroweak interactions including NP.
\end{itemize}

Clearly, there are other stars in flavour physics but I believe that the 
ones above will play the crucial role in our search for the theory of 
flavour. Having experimental results on these decays and observables with
sufficient precision accompanied by improved theoretical calculations will
exclude several presently studied models reducing thereby our exploration
of short distance scales to a few avenues.

\subsection{Superiority of Top-Down Approach in Flavour Physics}
Particle physicists are waiting eagerly for a solid evidence of NP for the
last 30 years. Except for neutrino masses, the BAU and dark matter, no clear 
signal emerged so far.
 While waiting for experimental signals several strategies for finding NP have been 
developed. In addition to  
precision calculations within the SM that allow to find   
the huge {\it background} to NP coming from
the known dynamics of this model (QCD corrections in 
flavour physics are reviewed in \cite{Buras:2011we}),
 one distinguishes between {\it bottom-up} 
and {\it top-down} approaches. Here I would like to express my personal 
view on these 
two approaches in the context of flavour physics and simultaneous 
 exploration of 
short distance physics both through LHC and high precision experiments.
\subsubsection{The Bottom-Up Approach}
In this approach one constructs effective field theories involving 
only light degrees 
of freedom including the top quark in which the structure of the effective 
Lagrangians is governed by the symmetries of the SM and often other 
hypothetical symmetries. This approach is rather powerful in the case of
electroweak precision 
studies and definitely teaches us something about $\Delta F=2$ 
transitions. In particular lower bounds on NP scales depending on the 
Lorentz structure of operators involved can be derived from the data 
\cite{Bona:2007vi,Isidori:2010kg}.

However, except for the case of  MFV and closely related 
approaches based on flavour symmetries, the bottom-up approach ceases, 
in my view, to be useful in $\Delta F=1$ decays, 
because of very many operators that are allowed to appear
in the effective Lagrangians with coefficients that are basically 
unknown \cite{Buchmuller:1985jz,Grzadkowski:2010es}. In this 
approach then the correlations between various $\Delta F=2$ and $\Delta F=1$ 
observables in $K$, $D$, $B_d$ and $B_s$ systems are either not visible or 
very weak, again except MFV and closely related approaches. Moreover 
the correlations between flavour violation in low energy processes and 
flavour violation in high energy processes to be studied soon at the LHC 
are lost. Again MFV belongs to a few exceptions.
\subsubsection{The Top-Down Approach}
My personal view shared by some of my colleagues is that the top-down 
approach is more useful in flavour physics. Here one constructs first 
a specific model with heavy degrees of freedom. For high energy processes,
where the energy scales are of the order of the masses of heavy particles 
one can directly use this ``full theory'' to calculate various processes 
in terms of the fundamental parameters of a given theory. For low energy 
processes one again constructs the low energy theory by integrating out 
heavy particles. The advantage over the previous approach is that now the 
coefficients of the resulting local operators are calculable in terms of 
the fundamental parameters of this theory. In this manner correlations between 
various observables belonging to different mesonic systems and correlations 
between low energy and high-energy observables are possible. Such correlations 
are less sensitive to free parameters than separate observables and 
represent patterns of flavour violation characteristic for a given theory. 
These correlations can in some models differ strikingly from the ones of 
the SM and of the MFV approach.

\subsection{Anatomies of explicit models}
\begin{figure}[hb]
\centerline{\includegraphics[width=0.65\textwidth]{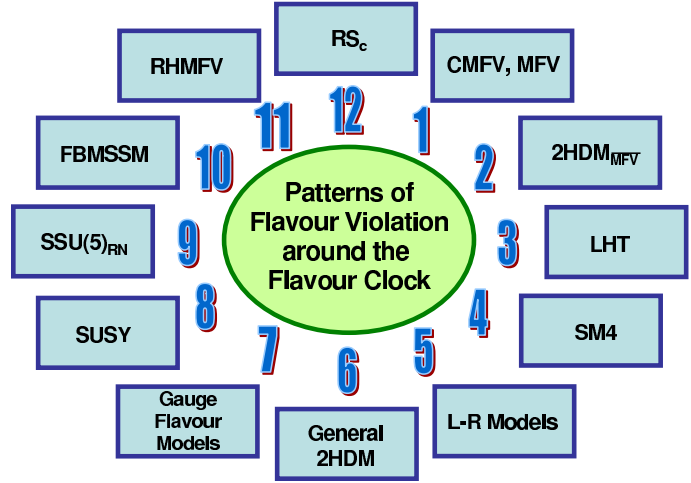}}
\caption{Various patterns of flavour violation around the Flavour Clock.}\label{Fig:2}
\end{figure}

Having the latter strategy in mind my group at the Technical University Munich, 
consisting dominantly of diploma students, PhD students and young post--docs 
investigated in the last decade flavour violating and CP-violating processes 
in a multitude of models. The names of models analyzed by us are collected in Fig.~\ref{Fig:2}. A summary of these studies before 2011 with brief descriptions of all these models can be found in  \cite{Buras:2010wr}. Below, I will 
frequently refer to these results and will briefly mention the most recent 
results obtained in my group.

\section{First Messages from New Animalcula}
\subsection{Preliminaries}
While the CKM+GIM picture of flavour and CP Violation describes the 
existing data surprisingly well, a number of anomalies observed 
in last years indicate that new Animalcula could be already in sight.
Three of these
 anomalies concern CP violation in $K_L$, $B_d$ and $B_s$ systems and 
all three are related to particle-antiparticle mixing. Also the data 
on the decay $B^+\to \tau^+\nu$ seem to depart from the SM expectations.
Last but certainly not least the anomalous magnetic moment of the muon, 
$(g-2)_\mu$, is by $3.2~\sigma$ above the SM value. Let me then briefly 
summarize these anomalies.

\subsection{The $\varepsilon_K-S_{\psi K_S}$ Anomaly}
It has been pointed out in \cite{Buras:2008nn,Buras:2009pj} that the SM prediction for $\varepsilon_K$ 
implied by the measured value of $S_{\psi K_S}=\sin 2\beta$, the ratio 
$\Delta M_d/\Delta M_s$ and the value of $|V_{cb}|$ turns out to be too
small to agree well with experiment. This tension between $\varepsilon_K$ and
 $S_{\psi K_S}$ has been pointed out from a different perspective in
\cite{Lunghi:2008aa,Lunghi:2009sm,Lunghi:2009ke,Lunghi:2010gv}.
These findings have been confirmed by a 
UTfitters  analysis \cite{UTfit-web}. 
The CKMfitters having a different treatment of uncertainties find less significant effects in $\varepsilon_K$ \cite{Lenz:2010gu}.

 Indeed taking the experimental value of $S_{\psi K_S}=0.672\pm 0.023$, 
$\vcb=0.0416$, the most
 recent value of 
the relevant non-perturbative 
parameter $\hat B_K=0.724\pm0.008\pm0.028$ \cite{Antonio:2007pb} 
(see also the most recent message from RBC and UKQCD collaborations 
\cite{Aoki:2010pe} 
$\hat B_K=0.749\pm0.027$) 
resulting from unquenched lattice 
calculations and including
long distance (LD) effects in 
${\rm Im}\Gamma_{12}$ and ${\rm Im}M_{12}$  in the $K^0-\bar K^0$ mixing
\cite{Buras:2008nn,Buras:2010pza} as well as recently calculated NNLO 
QCD corrections to $\varepsilon_K$ \cite{Brod:2010mj} one finds
\cite{Brod:2010mj}
\begin{equation}\label{epnew}
|\varepsilon_K|_{\rm SM}=(1.90\pm0.26)\cdot 10^{-3},
\end{equation}
visibly below the experimental value 
$|\varepsilon_K|_{\rm exp}=(2.23\pm0.01)\cdot 10^{-3}$. 

On the other hand
$\sin 2\beta=0.85\pm 0.05$ from SM  fits of the Unitarity Triangle (UT) 
is significantly larger
than the experimental value $S_{\psi K_S}=0.672\pm 0.023$. This discrepancy 
is  to some extent 
caused by the desire to fit $\varepsilon_K$ \cite{Lunghi:2008aa,Buras:2008nn,Buras:2009pj,Lunghi:2009sm,Lunghi:2009ke,Lunghi:2010gv} and 
$Br(B^+\to\tau^+\nu_\tau)$ \cite{Lunghi:2010gv}. For the most recent 
discussions including up to date numerics see 
\cite{Lenz:2011ti,Laiho:2011nz,Barbieri:2011ci}.

One should also recall the tension between inclusive and exclusive determinations of $|V_{ub}|$ with the exclusive ones in the ballpark of $3.5\cdot 10^{-3}$ and the 
inclusive ones typically above $4.0\cdot 10^{-3}$. 
As discussed in \cite{Crivellin:2009sd} an interesting solution to 
this problem is the presence of RH charged currents, which selects the inclusive value as the true value, implying again $\sin 2\beta\approx 0.80$ 
\cite{Buras:2010pz}.

As discussed in \cite{Lunghi:2008aa,Buras:2008nn} and subsequent papers of these authors a negative NP phase $\varphi_{B_d}$ in 
$B^0_d-\bar B^0_d$ mixing would solve both problems, provided such a phase 
is allowed by other constraints.  Indeed we have 
then 
\begin{equation}
S_{\psi K_S}(B_d) = \sin(2\beta+2\varphi_{B_d})\,, \qquad
S_{\psi\phi}(B_s) =  \sin(2|\beta_s|-2\varphi_{B_s})\,,
\label{eq:basic}
\end{equation}
where the corresponding formula for $S_{\psi\phi}$ in the presence of 
a NP phase $\varphi_{B_s}$ in 
$B^0_s-\bar B^0_s$ mixing has also been given. With a negative $\varphi_{B_d}$ 
the true $\sin 2\beta$ is larger than $S_{\psi K_S}$, implying a higher value
on $|\varepsilon_K|$, in reasonable agreement with data and a better UT-fit. This 
solution would favour the inclusive value of $|V_{ub}|$ as chosen 
e.g. by RH currents but as pointed out in \cite{Buras:2010pz} this 
particular solution of the ''$V_{ub}-problem''$ does not allow for a good 
fit to $S_{\psi K_S}$ if large $S_{\psi\phi}$ is required.
\subsection{Facing an enhanced CPV in the $B_s$ mixing}
The first detailed SM and model independent studies of CP violation in the 
$B^0_s-\bar B^0_s$ mixing in relation to Tevatron and LHCb experiments 
go back to \cite{Dighe:1998vk,Dunietz:2000cr}. Among the recent studies of 
these authors let me just quote \cite{Lenz:2011ti,Faller:2008gt}.

This topic became rather hot recently.
Indeed possibly the most important highlight in flavour physics in 2008, 2009 
\cite{Aaltonen:2007he} and even 
more in 2010 was the enhanced value of $S_{\psi\phi}$ measured by the CDF and 
D0 collaborations, seen either directly or indirectly through the 
correlations with various semi-leptonic asymmetries. While in 2009 and in 
the Spring of 2010  \cite{Abazov:2010hv}, the messages from Fermilab indicated good prospects 
for $S_{\psi\phi}$ above 0.5, the messages from ICHEP 2010 in Paris, 
softened such hopes a bit \cite{Aaltonen:2010xx}.
Both CDF and D0 find the enhancement 
by only one $\sigma$. Yet, this does not yet preclude $S_{\psi\phi}$ above 0.5, 
which would really be a fantastic signal of NP.
Indeed various recent fits indicate that $S_{\psi\phi}$ could be even as high 
as $0.8$ \cite{Lenz:2011ti}.
Let us hope that the future data from Tevatron and in 
particular from the LHCb, will measure this asymmetry with sufficient 
precision so that we will know to which extent NP is at work here. 
One should also hope that the large CPV in dimuon CP asymmetry from D0,
  that triggered 
new activities, will be better understood. I have nothing to add here
at present and can only refer to numerous papers  
\cite{Dobrescu:2010rh,Ligeti:2010ia,Blum:2010mj,Lenz:2010gu,Bauer:2010dg}.

In what follows I will decribe how different NP scenarios would face 
a future measurement of a significantly enhanced value of
$S_{\psi\phi}$.

The value $S_{\psi\phi}\ge 0.5$
can be obtained in the RSc model due to KK gluon 
exchanges and also heavy neutral KK electroweak gauge boson exchanges \cite{Blanke:2008zb}. See also \cite{Bauer:2009cf}.
In the  supersymmetric flavour models with the dominance of RH
currents  double Higgs penguins constitute the dominant NP 
contributions responsible for $S_{\psi\phi}\ge 0.5$, while in models where NP LH current 
contributions are equally important, also gluino boxes are relevant.
On the operator level LR operators 
are primarly responsible for this enhancement. Detailed analysis of this 
different cases can be found in \cite{Altmannshofer:2009ne}.
Interestingly the SM4 having only $(V-A)\times (V-A)$ operator  is also capable 
in obtaining 
high values of $S_{\psi\phi}$ \cite{Hou:2005yb,Soni:2008bc,Buras:2010pi,Soni:2010xh,Eberhardt:2010bm}.
In the LHT model where only $(V-A)\times (V-A)$ operators 
are present 
and the NP enters at higher scales than in the SM4, $S_{\psi\phi}$ above 
0.5 is out of reach \cite{Blanke:2009am}. 

All these models contain new sources of flavour  and CP violation 
and it is not surprising that in view of many parameters involved, large 
values of $S_{\psi\phi}$ can be obtained. The question then arises whether 
strongly
enhanced values of this asymmetry would uniquely imply new sources of 
flavour violation beyond the MFV hypothesis. The answer to this question is as 
follows:
\begin{itemize}
\item
In models with MFV and flavour blind phases (FBPs) set to zero, $S_{\psi\phi}$
remains indeed SM-like.
\item
In supersymmetric models with MFV even in the presence of 
 non-vanishing FBPs, at both 
small and 
large $\tan\beta$, the supersymmetry constraints do not allow values 
of $S_{\psi\phi}$ visibly different from the SM value 
\cite{Altmannshofer:2009ne,Blum:2010mj,Altmannshofer:2008hc}.
\item
In the ${\rm 2HDM_{\overline{MFV}}}$ in which at one-loop both Higgs
doublets couple to up- and down-quarks in the context of MFV, it is possible
 to obtain $S_{\psi\phi}\ge 0.5$ while satisfying 
all existing constraints \cite{Buras:2010mh}.
\end{itemize}

The driving force for large values of  $S_{\psi\phi}$ in this NP scenario 
are FBPs in interplay with the CKM matrix.\footnote{Various recent 
papers on FBPs not discussed here are collected in \cite{FBPs}. 
See, in particular, in the context of the Aligned two-Higgs-doublet model 
\cite{Pich:2009sp}. Also numerous studies of 2HDM models 
have been done by Gustavo 
Branco and his group. See \cite{Botella:2009pq} and earlier papers.} Dependently whether these 
phases appear in Yukawa couplings and/or Higgs potential one can distinguish 
three scenarios:

{\bf A)}
The FBPs in the Yukawa interactions are the dominant source 
of new CPV. In this case the NP phases 
 $\varphi_{B_s}$ and $\varphi_{B_d}$ are related through \cite{Buras:2010mh}
\begin{equation}\label{BCGI}
 \varphi_{B_d}\approx\frac{m_d}{m_s}\varphi_{B_s}\approx \frac{1}{17} \varphi_{B_s}.
\end{equation}
Thus in this scenario large $\varphi_{B_s}$ required to obtain values of 
$S_{\psi\phi}$ above 0.5 imply a unique small shift in $S_{\psi K_S}$ 
that allows to lower $S_{\psi K_S}$ from 0.74 down to 0.70, 
that is closer to the experimental value $0.672\pm0.023$. This in turn 
implies that it is $\sin 2\beta=0.74$ \footnote{The present value value from 
the most recent UT fit analyses mentioned above is a bit higher and close to 0.80 modifying a bit 
the numerics below.} and not  $S_{\psi K_S}=0.67$ that 
should be used in calculating $\varepsilon_K$ resulting in a value of 
 $\varepsilon_K\approx 2.0\cdot 10^{-3}$ within one $\sigma$ from the 
experimental value. The direct Higgs contribution to $\varepsilon_K$ 
is negligible because of small masses $m_{d,s}$. We should emphasize that 
once  $\varphi_{B_s}$ is determined from the data on $S_{\psi\phi}$ by means 
of (\ref{eq:basic}), the implications for $\varepsilon_K$ and  $S_{\psi K_S}$ are 
unique. 
The plots of $\varepsilon_K$ and  $S_{\psi K_S}$ 
versus $S_{\psi\phi}$ in \cite{Buras:2010mh} show this very transparently.
On the other hand this scenario does not provide any clue for the difference between inclusive and exclusive determinations of $|V_{ub}|$. Moreover, it appears 
that (see below) that the effect of FBPs in Yukawa couplings in a MFV framework 
is a bit to weak to solve quantitatively existing tensions.

{\bf B)}
The FBPs in the 
Higgs potential are are the dominant source 
of new CPV. In this case the NP phases 
 $\varphi_{B_s}$ and $\varphi_{B_d}$ are related through \cite{Ligeti:2010ia,Blum:2010mj} \footnote{This relation has been postulated already in 
\cite{Ball:2006xx,Buras:2008nn}.}
\begin{equation}\label{BG}
 \varphi_{B_d}=\varphi_{B_s}
\end{equation}
and the plots of $\varepsilon_K$ and  $S_{\psi K_S}$
versus $S_{\psi\phi}$ are strikingly modified \cite{Buras:2010zm}: the dependence is much 
stronger and even moderate values of $S_{\psi\phi}$ can solve all tensions. 
However, large values of $S_{\psi\phi}$ are not allowed if one wants to reproduce
the experimental value of $S_{\psi K_S}$.

{\bf C)} Hybrid scenario in which FBPs are present in both Yukawa interactions 
and Higgs potential so that \cite{Buras:2010zm}
\begin{equation}\label{BIP}
 \varphi_{B_d}=a \frac{m_d}{m_s}\varphi_{B_s}+ b\varphi_{B_s}=\kappa \varphi_{B_s},
\end{equation}
where $a,b,\kappa$ are real coefficients.

Presently it is not clear which relation between $\varphi_{B_s}$ and 
$\varphi_{B_d}$ fits best the data but the model independent analysis 
of \cite{Ligeti:2010ia} indicates that $\kappa\approx 1/5$.
Which of the two flavour-blind CPV mechanisms dominates depends on the value of
$S_{\psi\phi}$, which is still affected by a sizable experimental error, and
 also by the precise amount of NP allowed in $S_{\psi K_S}$.

\subsection{Implications of an enhanced  $S_{\psi\phi}$}
There are many implications of an enhanced value of  $S_{\psi\phi}$ in 
concrete NP models, which have been worked out in our papers. We 
have reviewed these implications in some details in   \cite{Buras:2010wr}. 
Here we will just collect some of the striking implications:
\begin{itemize}
\item
Enhanced $\Br(B_{s}\to \mu^+\mu^-)$ in SUSY flavour models,  ${\rm 2HDM_{\overline{MFV}}}$ and  SM4,
\item
Enhanced $\Br(B_{d}\to \mu^+\mu^-)$ in ${\rm 2HDM_{\overline{MFV}}}$ and in 
some SUSY flavour models,
\item
$\Br(B_{d}\to \mu^+\mu^-)$ forced to be SM-like in SM4,
\item
$\Br(K^+\to\pi^+\nu\bar\nu)$ and $\Br(K_L\to\pi^0\nu\bar\nu)$  forced 
to be SM-like in LHT \cite{Blanke:2009am} and RSc models 
\cite{Blanke:2008yr} but not in SM4.
\item
Automatic enhancements of $\Br(\mu\to e\gamma)$, $\Br(\tau\to \mu\gamma)$, 
$(g-2)_\mu$ and of EDMs $d_e$ and $d_n$ in SUSY-GUT models 
\cite{Altmannshofer:2008hc,Buras:2010pm}
\end{itemize}

We observe that simultaneous consideration of $S_{\psi\phi}$ and 
 $\Br(B_{s,d}\to \mu^+\mu^-)$ can already help us in eliminating
some NP scenarios. Even more insight will be gained when 
$\Br(K^+\to\pi^+\nu\bar\nu)$ and $\Br(K_L\to\pi^0\nu\bar\nu)$ will be 
measured. In particular if $S_{\psi\phi}$ will turn out to be SM-like
the branching ratios $\Br(K^+\to\pi^+\nu\bar\nu)$ and $\Br(K_L\to\pi^0\nu\bar\nu)$
can now be strongly enhanced in the LHT model \cite{Blanke:2009am} and the
RSc model \cite{Blanke:2008yr,Bauer:2009cf} with  respect to 
the SM but this is not guaranteed.
These patterns of flavour violations demonstrate very clearly the power of 
flavour physics in distinguishing different NP scenarios.

\subsection{$B_s\to\mu^+\mu^-$ and $B_d\to\mu^+\mu^-$}
The branching ratios ${\rm Br}(B_{s,d}\to\mu\bar\mu)$  are very 
strongly suppressed in the SM:
\be\label{BRSM1}
{\rm Br}(B_d\to\mu^+\mu^-)_{\rm SM} = (1.0\pm 0.1)\times 10^{-10},
\ee
\be\label{BRSM2}
{\rm Br}(B_s\to\mu^+\mu^-)_{\rm SM} = (3.2\pm 0.2)\times 10^{-9}\,.
\ee
and satisfy in the SM and CMFV models the relation
\cite{Buras:2003td}  
\be\label{R1}
\frac{{\rm Br}(B_{s}\to\mu\bar\mu)}{{\rm Br}(B_{d}\to\mu\bar\mu)}
=\frac{\hat B_{d}}{\hat B_{s}}
\frac{\tau( B_{s})}{\tau( B_{d})} 
\frac{\Delta M_{s}}{\Delta M_{d}}.
\ee
It involves
only measurable quantities except for the ratio $\hat B_{s}/\hat B_{d}$
that is known already now from lattice calculations within $3\%$
\cite{Lattice}.  

The upper bounds on ${\rm Br}(B_d\to\mu^+\mu^-)$ 
from CDF, D0 and LHCb are still by an order of magnitude larger than the 
SM predictions but in the coming years LHCb should be able to bring 
this upper bounds down within a factor two from the SM predictions or 
to discover NP. In these studies the methodology developed recently
 in \cite{Fleischer:2010ay} and presented by Niels Tuning at this 
conference should be very useful. The branching 
ratios in question can be enhanced even by an order of magnitude 
in a number of NP scenarios and the relation in (\ref{R1}) can also 
be strongly violated \cite{Buras:2010wr}.

\subsection{EDMs, $(g-2)_\mu$ and $Br(\mu\to e\gamma)$}
While I was dominantly discussing quark physics and flavour violating 
processes, these three observables are also very interesting. 
They are governed by dipole operators but describe 
different physics as far as CP violation and flavour violation is concerned. 
EDMs are flavour conserving but CP-violating, $\mu\to e \gamma$ is CP-conserving but lepton flavour violating and finally $(g-2)_\mu$ is lepton flavour conserving and CP-conserving. A nice paper discussing all these observables 
simultaneously is \cite{Hisano:2009ae}.

In concrete models there exist correlations between these three observables 
of which EDMs and $\mu\to e\gamma$ are very strongly suppressed within the 
SM and have not been seen to date. $(g-2)_\mu$ on the other hand has been very precisely measured and exhibits a $3.2\sigma$ departure
 from the very precise SM value 
(see \cite{Prades:2009qp} and references therein) \footnote{ In a very 
recent paper it is claimed that the SM agrees perfectly with the 
data \cite{Bodenstein:2011qy}, confirmation or disproval of this claim 
would be very important.}.
Examples of these correlations can be found in 
\cite{Altmannshofer:2009ne,Altmannshofer:2008hc}. 
In certain supersymmetric 
flavour models with non-MFV interactions the solution of the $(g-2)_\mu$ 
anomaly implies simultaneously $d_e$ and $\Br(\mu\to e \gamma)$ in the reach of experiments in this decade. In these two papers several correlations 
of this type have been presented. 

The significant FBPs required to reproduce the enhanced value of $S_{\psi\phi}$ 
in the  ${\rm 2HDM_{\overline{MFV}}}$ model, necessarily  imply large EDMs 
 of the neutron, Thallium and Mercury atoms.
Yet, as a detailed
analysis in  \cite{Buras:2010zm} shows the present upper bounds on the
 EDMs do not forbid sizable non-standard CPV effects in $B_{s}$ mixing.
However, if a large CPV phase in $B_s$ mixing will be confirmed, this
will imply hadronic EDMs very close to their present experimental bounds,
within the reach of the next generation of experiments. For a recent model 
independent analysis of EDMs see \cite{Batell:2010qw}.

\subsection{Waiting for precise predictions of 
$\varepsilon'/\varepsilon$}
The flavour studies of the last decade have shown that provided the hadronic 
matrix elements of QCD-penguin and electroweak penguin operators will be 
known with sufficient precision, $\varepsilon'/\varepsilon$ will play a very 
important role in constraining NP models. We have witnessed recently an 
impressive progress in the lattice evaluation of $\hat B_K$ that elevated 
$\varepsilon_K$ to the group of observables relevant for precision studies 
of flavour physics. Hopefully  this could also 
be the case of  $\varepsilon'/\varepsilon$ already in this decade.

\subsection{$B^+\to \tau^+\nu_\tau$}
Another prominent anomaly in the data not discussed by us sofar is found in
the tree-level decay $B^+ \to \tau^+ \nu_\tau$. 
Within the SM we found
\cite{Altmannshofer:2009ne}
\begin{equation}\label{eq:BtaunuSM1}
{\Br}(B^+ \to \tau^+ \nu_\tau)_{\rm SM}= (0.80 \pm 0.12)\times 10^{-4},
\end{equation}
which 
agrees well with the result presented by the UTfit collaboration
\cite{Bona:2009cj}. 

On the other hand, the present experimental world avarage based 
on results by BaBar and Belle
reads \cite{Bona:2009cj}
\begin{equation} \label{eq:Btaunu_exp}
{\Br}(B^+ \to \tau^+ \nu_\tau)_{\rm exp} = (1.73 \pm 0.35) \times 10^{-4}~,
\end{equation}
which is roughly by a factor of 2 higher than the SM value.
We can talk about a tension at the $2.5\sigma$
level.

With a higher value of $|V_{ub}|$ as obtained through inclusive determination 
this discrepancy can be decreased significantly. For instance with a value 
of $4.4\times 10^{-3}$, the central value predicted for this branching 
ratio would be more like $1.25\times 10^{-4}$. Yet, this would then require 
NP phases in $B_d^0-\bar B_d^0$ mixing to agree with the data on $S_{\psi K_S}$. 
In any case values of ${\Br}(B^+ \to \tau^+ \nu)_{\rm exp}$ significantly 
above $1\times 10^{-4}$ will signal NP contributions either in this decay 
or somewhere else. For a very recent discussion of such correlations see 
\cite{Lunghi:2010gv}.

While the final data from BaBar and Belle will lower the exparimental
error on  $\Br(B^+\to\tau^+\nu)$, the full clarification of a possible
discrepancy between the SM and the data will have to wait for the
data from Belle II and SFF in Rome. Also improved values for $F_B$ from lattice 
and $\vub$ from tree level decays will be important if some NP like
charged Higgs is at work here. As a significant progress made by lattice 
groups \cite{Lattice} is continuing, there are good chances that around 2015 the picture of 
$B^+\to\tau^+\nu $ will be much clearer. The same applies to many $B$ 
physics observables as well.

\section{Messages from the last moment}
Finally, I would like to report on two recent papers from Munich.

In the first paper a minimal theory of fermion masses (MTFM) has been 
constructed \cite{Buras:2011ph}. This amounts to extend the SM by heavy vectorlike fermions 
with flavour-anarchical Yukawa couplings that mix with chiral fermions 
such that small SM Yukawa couplings arise from small mixing angles. This 
model can be regarded as an effective description of the fermionic 
sector of a large class of existing models and thus might serve as a 
useful reference frame for a further understanding of flavour hierarchies 
in the SM. Already such a minimal framework implies modifications in 
the couplings of $W^\pm$, $Z$ and $H$ to fermions and leading to novel 
FCNC effects with a special structure of their suppression that is different 
from MFV. This work shows once again 
that models 
attempting the explanation of the hierarchies of fermion masses and of 
its hierarchical flavour violating and CP violating interactions in 
most cases imply non-MFV interactions. This is also evident from the study 
of supersymmetric flavour models \cite{Altmannshofer:2009ne} and more general recent studies \cite{Lalak:2010bk,Dudas:2010yh}.
Further phenomenological implications of MTFM will be presented soon.

In the second paper \cite{Buras:2011zb} considering
a general scenario with new heavy neutral gauge bosons, present in particular
in $Z'$ and gauge flavour models, we have pointed out two new contributions to
the $\overline{B}\to X_s\gamma$ decay. The first one originates from one-loop diagrams mediated by gauge bosons and heavy exotic quarks with electric charge $-1/3$. The second contribution stems from the QCD mixing of neutral current-current operators generated by heavy neutral gauge bosons and the dipole operators responsible for the $\overline{B}\to X_s\gamma$ decay. The latter mixing is calculated 
in our paper for the 
first time.
We also  discussed general sum rules which have to be satisfied in any
model of this type. We emphasise that the 
neutral gauge bosons in question could also significantly affect other fermion radiative decays as well as non-leptonic two-body $B$ decays, $\epsilon'/\epsilon$, anomalous $(g-2)_\mu$ and electric dipole moments. Implications of these 
findings for concrete models will be presented soon. 

\section{Grand Summary}
I hope I 
convinced the readers that flavour physics is a very rich field which 
necessarily will be a prominent part of a future theory of fundamental 
interactions both at large and short distance scales. While MFV could 
work to first approximation, recent data indicate that at certain
level non-MFV interactions could be present.

What role will be played by flavour blind phases in future phenomenology 
depends on the future experimental data on EDMs. Similar comment applies 
to LFV. A discovery of $\mu\to e\gamma$ rate at the level of $10^{-13}$ 
would be a true mile stone in flavour physics. Also the discovery of 
$S_{\psi\phi}$ at the level of 0.3 or higher would have a very important 
impact on quark flavour physics. The measurements of 
$\Br(B_{s,d}\to \mu^+\mu^-)$ in conjunction with $S_{\psi\phi}$, $\kpn$ 
and at later stage $\klpn$ will allow to distinguish between various 
models. Here the correlations between 
various observables will be crucial. It is clearly important to clarify 
the origin of the tensions between $\varepsilon_K$, $S_{\psi K_S}$, $|V_{ub}|$ 
and $\Br(B^+\to \tau^+\nu_\tau)$ but this possibly has to wait until Belle II 
and later SFF will enter their operation.

In 
any case I have no doubts that we will have a lot of fun with flavour 
physics in this decade and that this field will offer very important 
insights into the short distance dynamics.

{\bf Acknowledgements}\\
I would like to thank the organizers of the Beauty 2011, in 
particular Robert Fleischer, for inviting me to such a pleasant 
conference.  I really enjoyed this stay in Amsterdam and the physics 
discussions we had. 
I also thank all my collaborators for exciting time we spent together 
exploring the short distance scales with the help of flavour violating 
processes.
This research was partially supported by the Cluster of Excellence `Origin and Structure
of the Universe' and  by the German `Bundesministerium f\"ur Bildung und Forschung'under contract 05H09WOE.



\end{document}